\documentclass[a4paper,twocolumn]{article}
\usepackage{graphicx}
\usepackage{amsmath}
\usepackage{amsfonts}
\usepackage{amssymb}

\begin{document}

\title{Comment on ''Giant Absorption Cross Section of Ultracold Neutrons in Gadolinium''}
\author{ }
\maketitle

Rauch \emph{et al }\cite{rauch1}\emph{\ }have measured the absorption cross
section of \ natural $Gd$ and isotopically enriched $^{157}Gd$ for neutron
energies extending into the ultracold energy region \cite{ucnbook} using $Gd$
compounds dissolved in $D_{2}O$. For the case of $^{157}Gd$ the result for
neutrons with a velocity of 10 \emph{m/sec} (49.7 \emph{Mbarns}) was found to
be less than the value obtained by extrapolating the value at thermal energies
($v=$ 2200 \emph{m/sec, }253,300 \emph{barns}) by the $1/v$ law (55.9
\emph{Mbarns}) (see Table 1 of \cite{rauch1}). Note that the $1/v$ law applies
to the velocity in the material. At low energies this differs from the
velocity in free space because of refraction effects. \cite{steyerl1}

The authors then attributed this discrepancy to the effect of random
fluctuations of the number of scattering centers in the interaction volume.
Because of the exponential nature of the absorption law, \ fluctuations to
lower densities have a greater effect and the observed transmission is larger
than it would be otherwise, leading to a reduction in the apparent cross
section. The authors suggest that the interaction volume is delimited by the
transverse 'coherence' lengths ($1/2\delta k)$ where $\delta k$ is the width
of the transverse momentum distribution, and longitudinally by the sample
thickness or absorption length, whichever is smaller.

The purpose of this note is twofold:

\begin{enumerate}
\item  To point out that the absorption cross section of $^{157}Gd$ has a
resonance in the thermal region as shown in fig.1 taken from \cite{ngatlas}.
The correct $1/v$ extrapolation of the data is thus seen to yield a value of
\ 42.5 \emph{Mbarns} so that the discrepancy (if any) with the measured value
has the opposite sense of that predicted by the model based on fluctuations in
an interaction volume.

\item  The transverse 'coherence' length is the width of the transverse
spatial correlation function of the wave function (we refer to it from now on
as the correlation length). In the usual case where the beam%
\begin{center}
\includegraphics[
height=3.32in,
width=3.55in
]%
{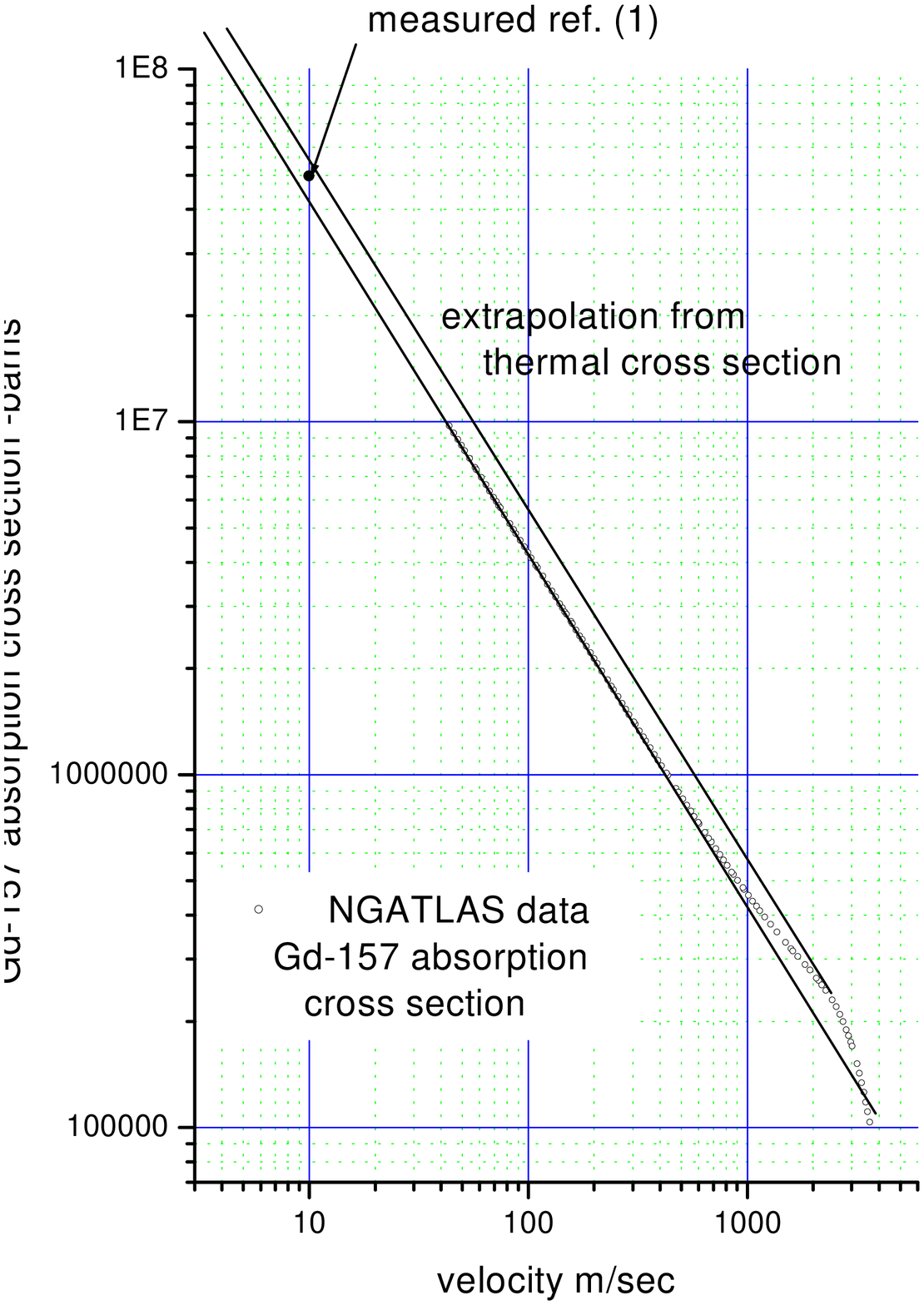}%
\\
Fig.1) Gd-157, tabulated absorption cross section data and extrapolations.
\end{center}
intensity varies slowly on the scale of the correlation length, the
correlation function is a measure of the average phase difference between
adjacent points, \ as shown in fig. 2 of \cite{sptime} and as such has no
influence at all on absorption, which depends only on the beam intensity.
Contrary to scattering which involves an interference between at least two
points in the sample \cite{sptime}, \cite{sinha} absorption takes place at a
single point and hence is not influenced by the correlation properties of the
beam. This can also be seen by applying the argument of Comsa \cite{comsa}.
\end{enumerate}

\bigskip

{\small J. Felber}$^{1}$, {\small R. G\"{a}hler}$^{1}$, {\small R. Golub}$^{2}$

$^{1}${\small Fakukt\"{a}t der Physik}

{\small \ Technische Universit\"{a}t M\"{u}nchen}

{\small \ 85748 Garching, Germany}

$^{2}${\small Hahn Meitner Institut}

{\small Glienickerst. 100, }

{\small 14109 Berlin, Germany}

\end{document}